\documentclass[prl,aps,amssymb,amsfonts,amsmath,showpacs,twocolumn]{revtex4-1}
\pdfoutput=1
\usepackage{graphicx}
\usepackage{ucs} 
\usepackage{color} 

\definecolor{darkblue}{rgb}{0,0,0.5}
\definecolor{lila}{rgb}{0.3,0,0.3}
\definecolor{turq}{rgb}{0,0.1,0.4}
\definecolor{lightblue}{rgb}{0.7,0.7,0.9}

\usepackage[pdftex,
  colorlinks=true,
  backref=page,
  linkcolor=darkblue, 
  filecolor=red,
  citecolor=turq, 
  urlcolor=lila, 
  pdftitle={Coherent Nonlinear Single Molecule Microscopy},
  pdfauthor={Ilja Gerhardt, Gert Wrigge, Jaseuk Hwang, Gert Zumofen, Vahid Sandoghdar},
  pdfsubject={Using the nonlinear response of a Single Molecule for Microscopic applications},
  pdfkeywords={single molecules, single photons source, coherent spectroscopy, short pulse, coherent microscopy, localization accuracy},
  pdfpagelabels=true, 
  breaklinks=false,
  plainpages=false,
  backref=false,
  bookmarks,
  bookmarksnumbered=true]{hyperref}

\newcommand{\textw}{8.5cm}
\newcommand{\beq}{\begin{eqnarray}}
\newcommand{\eeq}{\end{eqnarray}}

\newcommand{\Om}{{\mathit \Omega}}

\begin{document}

\title{Coherent Nonlinear Single Molecule Microscopy}

\author{I.~Gerhardt}\email{ilja@quantumlah.org}\altaffiliation{present address: 3.~Physikalisches Institut, Pfaffenwaldring~57, D-70550 Stuttgart, Germany}
\author{G.~Wrigge}
\author{J.~Hwang}
\author{G.~Zumofen}
\author{V.~Sandoghdar}
\affiliation{Laboratory of Physical Chemistry, ETH Zurich, CH-8093 Zurich, Switzerland}

\begin{abstract}

We investigate a nonlinear localization microscopy method based on
Rabi oscillations of single emitters. We demonstrate the fundamental
working principle of this new technique using a cryogenic far-field
experiment in which subwavelength features smaller than $\lambda$/10
are obtained. Using Monte Carlo simulations, we show the superior
localization accuracy of this method under realistic conditions and
a potential for higher acquisition speed or a lower number of
required photons as compared to conventional linear schemes. The
method can be adapted to other emitters than molecules and allows
for the localization of several emitters at different distances to a
single measurement pixel.
\end{abstract}

\pacs{42.50.Gy, 42.50.Nn, 42.62.Fi}
\maketitle

Optical microscopy has experienced many revolutionary developments
in the past two decades such as
scanning near-field optical microscopy
(SNOM)~\cite{pohl:84,lewis_ultramicroscopy_1984}, two-photon
confocal microscopy, coherent antistokes Raman scattering
(CARS)~\cite{PhysRevLett.82.4142}, stimulated emission depletion (STED)~\cite{hell_science_2007},
and single-molecule localization microscopy~\cite{Betzig:95,pertsinidis_nature_2010}. One of the
highest three-dimensional spatial resolution achieved so far has
reached about 2~nm~\cite{hettich_science_2002} and was based on the
latter concept, where the locations of individual molecules are
determined by finding the centers of their diffraction-limited
point-spread functions~\cite{bobroff:1152,thompson__2002}. The
requirement for this technique is the distinguishability of
neighboring molecules so that each point-spread function can be
examined separately. The first demonstration of this concept was at
low temperature~\cite{hettich_science_2002}, where the
inhomogeneous distribution of frequencies in the sample provides a
convenient ``spectral identity'' for each molecule. More recent
approaches have used stochastic photo-activation or switching
schemes for addressing single
molecules~\cite{Betzig2006,rust_natmeth_2006}. These
room-temperature experiments have caused a great deal of interest
for their applicability to biological systems, but their accuracy is
limited by the number of recorded photocycles before a single
molecule undergoes photobleaching.

In this article, we discuss a new scheme of nonlinear localization
microscopy based on the observation of coherent Rabi
oscillations~\cite{gerhardt_pra_2009}. In addition to a first experimental
demonstration, we present Monte-Carlo simulations and examine the
attainable localization accuracy as a function of the number of
recorded photons, pixel size, and number of Rabi oscillations. In
particular, we compare the performance of coherent Rabi imaging
microscopy (CORIM) with standard localization methods, where the
spatial distribution of the point-spread function depends linearly
on the signal intensity.

\begin{figure}[b!]
\centering
\includegraphics[width=7.0cm]{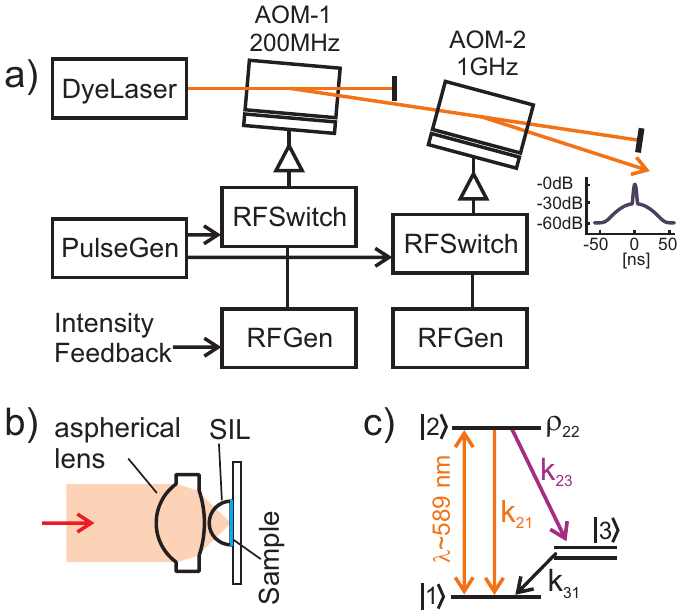}
\caption{(a) Experimental setup with two cascaded acousto-optical
modulators to obtain short laser pulses with a high signal to
background ratio. (b) The optical setup inside the cryostat with a
solid-immersion lens (SIL). (c) The level scheme of a single
molecule.\label{fig:setup}}
\end{figure}


\subsection{Experimental localization of a single molecule via position-dependent Rabi oscillations}

In this work we studied the dye molecule dibenzantanthrene (DBATT)
embedded in a {\emph n}-tetradecane Shpol'ski\unichar{0301} matrix.
The excited state in DBATT has a fluorescence lifetime of
$T_1=9.5$~ns, corresponding to a radiatively broadened linewidth of
$\mathit{\Gamma}_1/2\pi= 17$~MHz for the zero phonon line (ZPL)
transition. We used a CW single-mode dye laser (Coherent 899
Autoscan, $\Delta \nu \leq 1\,{\rm MHz}$) to identify single
molecules via fluorescence excitation spectroscopy~\cite{Orrit:90}.
For all further experiments we chopped the beam by means of two
cascaded acousto optical modulators (AOMs), achieving pulse width
down to 2.9~ns and a signal to background ratio of more than 60~dB.
Further details of the pulse generation scheme are explained in a
previous publication~\cite{gerhardt_pra_2009}. As sketched in
Fig.~\ref{fig:setup}b, a confocal microscope based on a
solid-immersion lens enabled us to efficiently excite a molecule and
detect it both via its Stokes-shifted fluorescence and via
extinction spectroscopy~\cite{Wrigge:08}. The red-shifted
fluorescence was filtered from the excitation laser light by an
optical long pass filter and was detected using an avalanche
photodiode (APD).

\begin{figure}[t!]
\centering
\includegraphics[width=7.0cm]{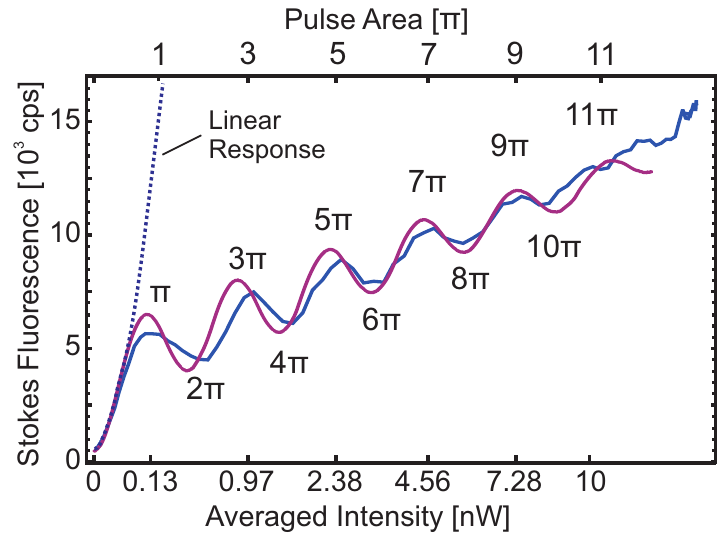}
\caption{Intensity dependence of the response from a single molecule
in temporally integrated detection. Pulse duration was 4~ns,
repetition rate was 700~kHz, and the intensity was integrated over
100~ms. Figure adapted
from~\cite{gerhardt_pra_2009}.\label{fig:intensityresponse}}
\end{figure}

In coherent state preparation, the system can undergo several Rabi
cycles, depending on the pulse area ($A$) which is proportional to
the square root of the intensity and the duration of the excitation
pulse. Figure~\ref{fig:intensityresponse} shows the experimental
results for the integrated Stokes-shifted optical response of a
single molecule excited by a series of 4~ns long narrow-band pulses.
The signal was averaged for 100~ms and is plotted against the
excitation intensity, depicted on a quadratic scale. We note in
passing that as illustrated by the dotted line in
Fig.~\ref{fig:intensityresponse}, a purely linear response as used
in conventional localization techniques would appear as a quadratic
function in this plot.

\begin{figure}[tb]
\centering
\includegraphics[width=7.0cm]{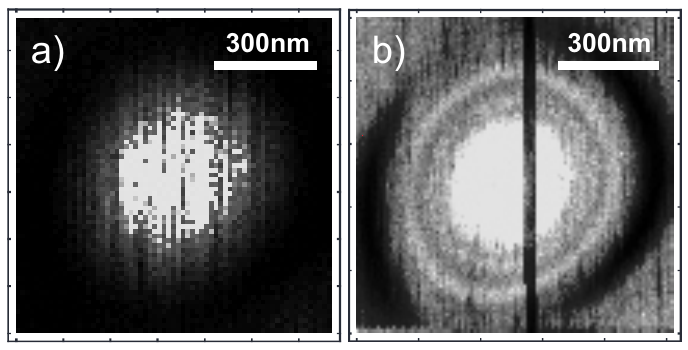}
\caption{Experimental raster-scanned images of a single molecule
without (a) and with (b) pulsed excitation (4~ns pulses, 700~kHz
repetition rate). The effectively larger point-spread function of
the illumination spot due to the field dependence is visible. The
rings around the central spot have widths of $\approx$ 40~nm.
\label{fig:experimental}}
\end{figure}

If the focal spot of the excitation beam is scanned over the sample,
the single emitter experiences different values of $A$ at various
lateral coordinates, and the emitted fluorescence signal appears as
several concentric rings. The width of each ring depends on the
gradient of the intensity at that point and is the smallest at the
highest slope of the optical point-spread function. The center of
the acquired image depends on the maximum excitation intensity of
the molecule and corresponds to several excitation and de-excitation
cycles of the emitter. Figures~\ref{fig:experimental}a and b show
the experimental results of such a measurement on a single molecule
with continuous-wave (CW) and pulsed excitations, respectively. The
occasional dark regions in the CW image reveal typical triplet-state
blinking, while the full width at half-maximum (FWHM) of 370~nm
indicates an almost diffraction-limited focusing spot.

The raster-scanned image with pulsed excitation displays higher
spatial frequencies and spatial features down to $\approx$~40~nm.
The dark stripe in the middle of the image was caused by a temporary
spectral jump of the molecular resonance and provides a convenient
measure for the background fluorescence. We also note the asymmetry
of the images in Fig.~\ref{fig:experimental}, which might be caused
by a cushion distortion slightly away from the optical axis of the
solid-immersion lens~\cite{ippolito:053105}.

\subsection{Analytical considerations}
\begin{figure}[t,b]
\centering
\includegraphics[width=7.0cm]{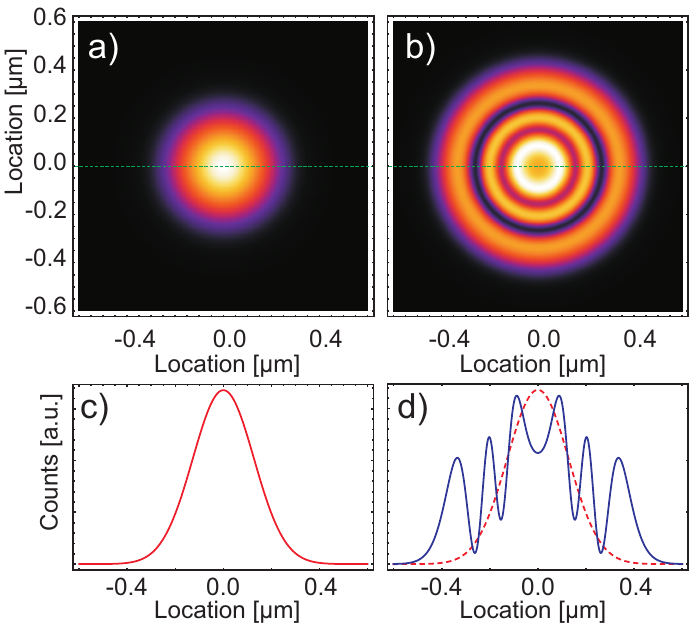}
\caption{Simulated raster-scanned image of a single molecule and
line scans. (a) The usual lineear optical response of a single
molecule. (b) If the excitation light is pulsed and the pulse width
is shorter than the singlet $T_{\rm 1}$ time of the molecule,
several rings appear, representing Rabi oscillations. c, d) Cross
sections of (a) and (b). \label{fig:simul}}
\end{figure}

In Fig.~\ref{fig:simul}a, we first consider the simulated linear
response $\widetilde{P}_{red} \propto I_{laser}$ of a single
molecule, where $P_{red}$ denotes the red-shifted fluorescence
signal. One can localize the molecule as is common in conventional
localization schemes by applying a numerical regression to an
assumed Gaussian intensity
distribution~\cite{bobroff:1152,thompson__2002}. The highest spatial
frequency in this imaging configuration is given by the
diffraction-limited profile of the laser beam focus.

We now consider coherent excitation with light pulses short compared
to the lifetime, $t_p \ll T_1$. Under this condition, the molecular
polarization and the population of the ground and excited states
oscillate at the Rabi frequency
\beq
\Om = d E(x_0,y_0)/\hbar
\eeq
where $E(x_0,y_0)$ is the electric laser field at the position
$\{x_0,y_0\}$ of the molecule and $d$ is the transition dipole
moment in the direction of the field. The state at the end of the
pulse is determined by the pulse area which for a rectangular pulse
shape reads
\beq
A = \Om t_p \propto \sqrt{I_{laser}}.
\eeq As a result, the excited-state population at the end of the
pulse follows a sinusoidal shape with increasing $A$. Accordingly,
the emission probability of a photon per pulse is given by
\beq
P = \left( \sin A /2 \right )^2.
\label{eqn:nobg}
\eeq
Assuming a Gaussian focal shape, the electric field reads
\beq
E(x_0,y_0) = E_0 e^{ - \left[(x-x_0)^2+(y-y_0)^2\right]/ (2 w_0^2) }
\eeq where $\{x,y\}$ is the position of the focal spot which is
scanned over the $x$-$y$ plane. The parameter $w_0$ gives the
characteristic width of the Gaussian shape. Thus, finds the
photon-emission probability to be
\beq
P =  \sin^2 (f \sqrt{ I_0}
e^{ - \left[(x-x_0)^2+(y-y_0)^2\right]/(2 w_0^2) })
\label{eqn:psf}
\eeq
where $I_0$ is the light intensity at the focal spot and $f$ is
a scaling parameter. In what follows, we use this expression for the
Monte-Carlo simulations.

Experimental measurements such as those in
Fig.~\ref{fig:experimental} might deviate from the ideal response.
Usually the observed Rabi oscillations might be damped for three
reasons: a)~a finite pulse length and the dephasing time $T_2$
reduce the modulation depth below unity, b)~due to the finite pulse
length, the molecule decays and is re-excited within the pulse
duration several times, and c)~the pulse area fluctuations wash out
the visibility of the integrated signal. These effects are
collectively included in the following equations by damping the
oscillations with a field-dependent term $a$. Furthermore, we add a
linear contribution $b$ to account for the background, which scales
roughly linearly with the optical field as experimentally observed
in~\cite{gerhardt_pra_2009}. Also other molecules might spectrally
overlap and lead to a background contribution which is then
proportional to the excitation intensity, modeled by $c$. Moreover,
the intensity-independent pixel noise (i.e.~dark counts from a
photodetector) can be denoted by $r$. With the above-mentioned
parameters it is possible to describe the full optical response in
the following form \beq
   I_d = \eta \frac{  \sin^2  \left( f \sqrt{I} \right)}{a \sqrt{ I}}
     + b \sqrt I +  c I + r
\label{eq6} \eeq where $\eta$ covers the collection efficiency,
pulse repetition rate, and the reduction by filters used to
discriminate between laser and fluorescence light.
Figures~\ref{fig:simul}b and d show the contributions of the first
and second terms on the right-hand side of Eqn.~(\ref{eq6}).

\subsection{Monte Carlo analysis of the localization accuracy}
\begin{figure}[thb!]
\centering
\includegraphics[width=7.0cm]{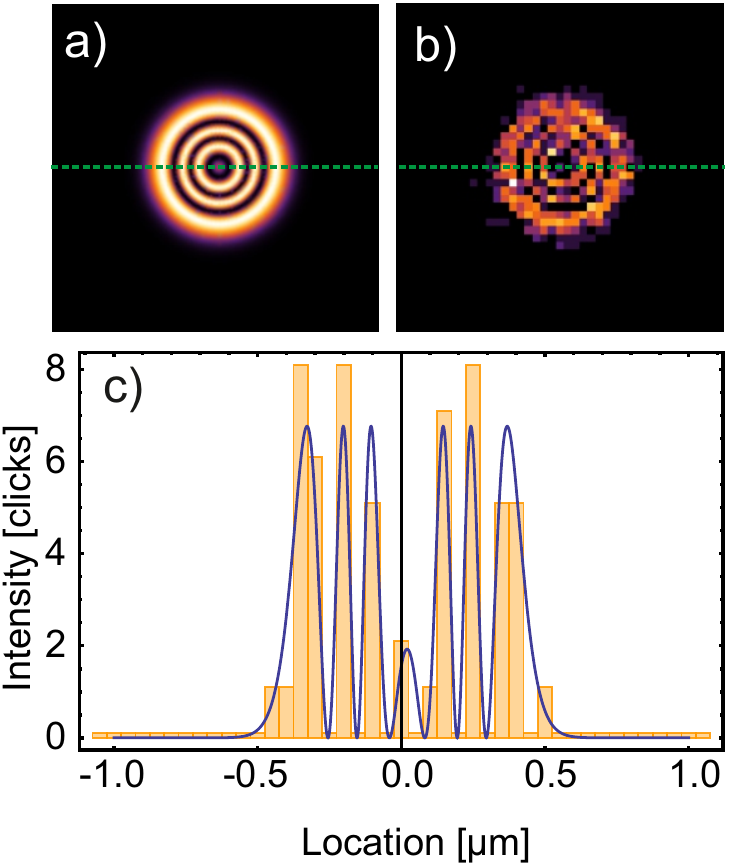}\\
\caption{Basic derivation of pixilated images for Monte Carlo
estimations: (a) Calculated point-spread function in the absence of
amplitude damping and detector noise (analytical).  (b) Numerically
estimated pixel distribution for 1000 detected photons. (c) Line
cuts for pixel distribution and fitted point-spread
function.\label{fig:fitexplanations}}
\end{figure}

In order to assess the promise of CORIM for localization microscopy,
we performed Monte Carlo simulations using \texttt{Mathematica 7.01}
(Wolfram Research). For realistic experimental parameters, we used a
statistical evaluation by generating images followed by a numerical
fitting procedure. This method allows us to judge the localization
accuracy for a variety of parameters such as the number of detected
photons. For each data point and set of parameters, a series of 500
images was generated, by filling a grid of the defined pixel size
according to the distribution function (see
Fig.~\ref{fig:fitexplanations}a) given by Eqn.~(\ref{eqn:psf}) until
the detected number of photons was reached.
Figure~\ref{fig:fitexplanations}b displays an example of the
Monte-Carlo image. The center of the emitter was randomly chosen to
be within the inner 2$\times$2 pixel. The pixel grid was defined
such that the range around the molecule was larger than 1~$\mu$m by
one pixel. The resulting image was fitted by the point-spread
function using the \texttt{Mathematica} internal function for
nonlinear fitting without any further constraints
(\texttt{NonlinearModelFit}). The cross section in
Fig.~\ref{fig:fitexplanations}c shows an example. Here, experimental
background noise has not been included so that only pixilation and
shot noise contribute to the observed fluctuations.

The start parameters of the simulations that follow below included
the actual center of the spot, the height of the highest pixel as an
amplitude, and $w_0$ of the optical point-spread function ($w_0 =
\frac{1}{2}\, \mbox{FWHM}/\sqrt{2 \ln{2}} \approx$~120~nm). The 500
Monte Carlo runs were unweighted averaged. The confidence range
($\pm 1~\sigma$) of one coordinate is displayed. Unless differently
specified, the parameters were as follows: Gaussian focus,
FWHM=300~nm, detection of 200 photons, pixel size of 50~nm, and a
Rabi parameter of 100, leading to approximately 3 Rabi flops
(6.2~$\pi$). These parameters are labeled by the dashed lines in the
presented figures.

\begin{figure}[thb!]
\centering
\includegraphics[width=\textw]{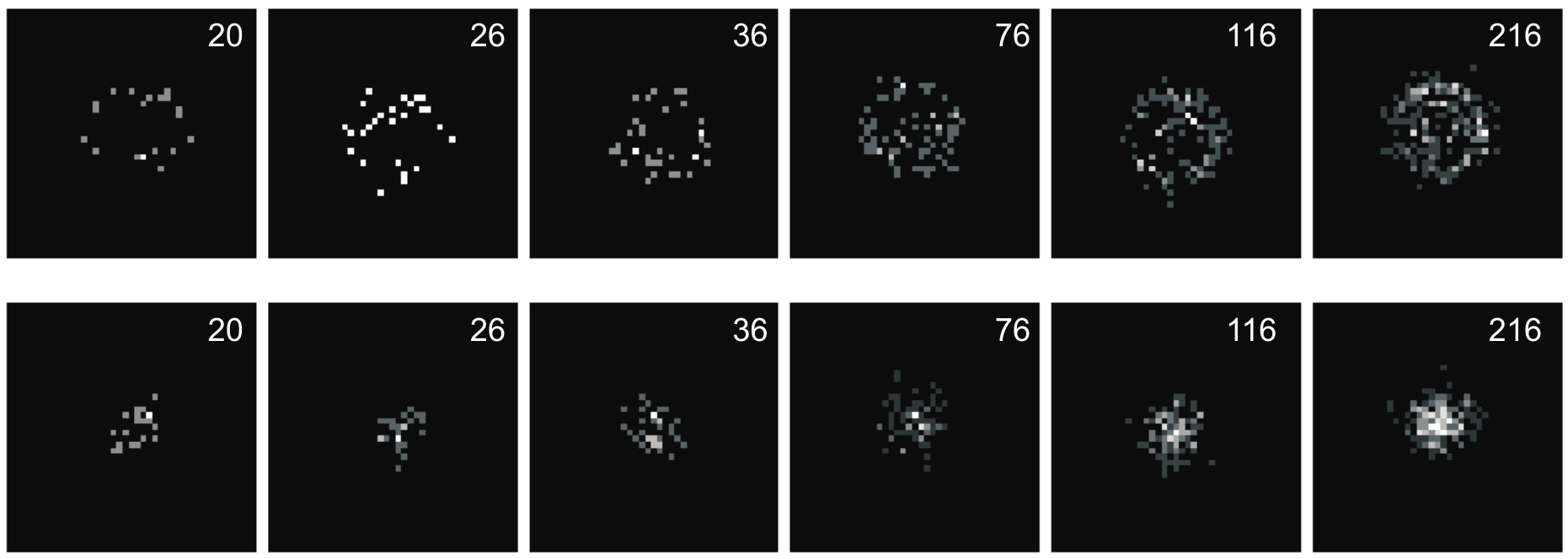}\\
\includegraphics[width=7.0cm]{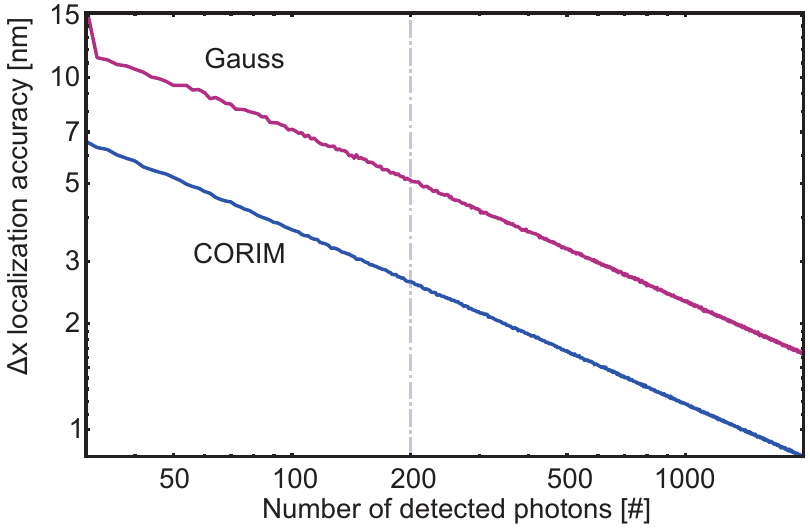}
\caption{Localization accuracy for different numbers of detected
photons. Gaussian fitting (upper red curve) and coherent Rabi
imaging microscopy (lower blue curve). Both curves scale as
$1/\sqrt{N}$, as expected for pure shot-noise limited localization
performance (background noise is neglected in these calculations).
For details of this simulation see text.\label{fig:mc02}}
\end{figure}

The localization accuracy with $N$ detected photons scales as
1/$\sqrt{N}$, depending on the shot noise of the measurement. This
behavior is observed both for linear imaging and CORIM. At the top
of Fig.~\ref{fig:mc02}, we show CORIM and Gaussian simulations for
different $N$. The bottom part of Fig.~\ref{fig:mc02} summarizes the
results and compares the localization error between the two cases.
We find that the accuracy in CORIM is superior by a factor of two
for the given parameters. Thus, a single emitter can be localized
with higher accuracy in a shorter time, which might offer a crucial
advantage against systematic errors caused by mechanical drift and
jitter~\cite{bloess01}.

\begin{figure}[thb!]
\centering
\includegraphics[width=\textw]{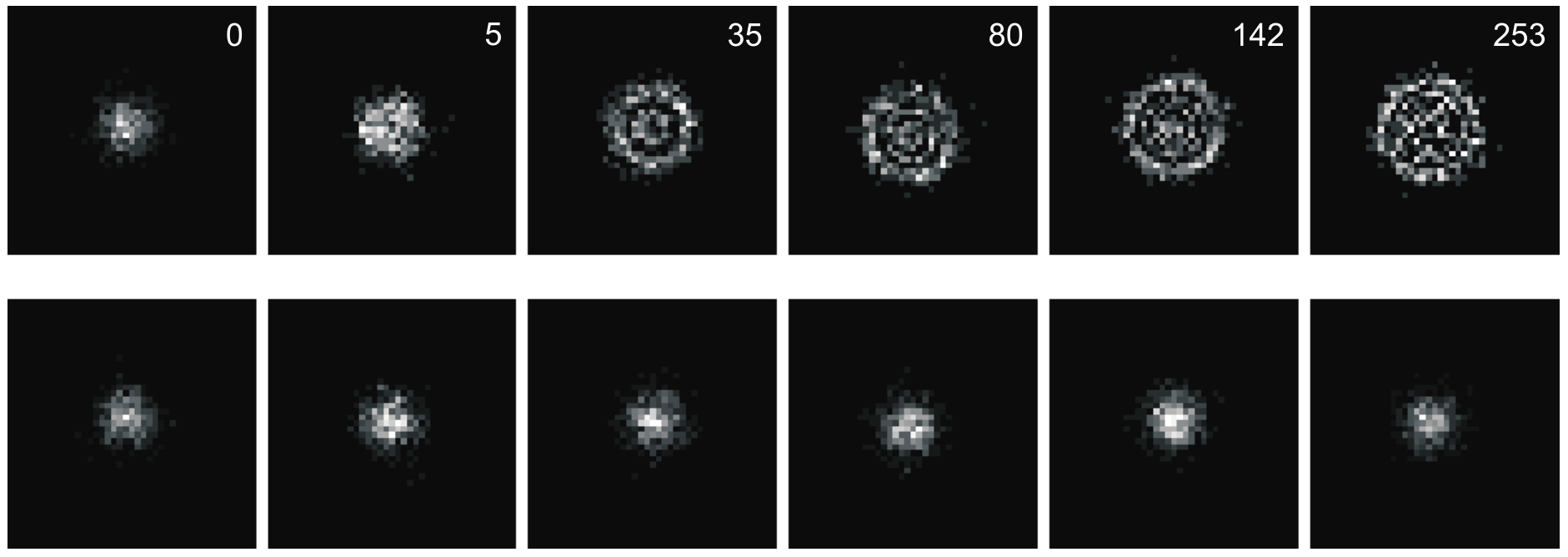}\\
\includegraphics[width=7.0cm]{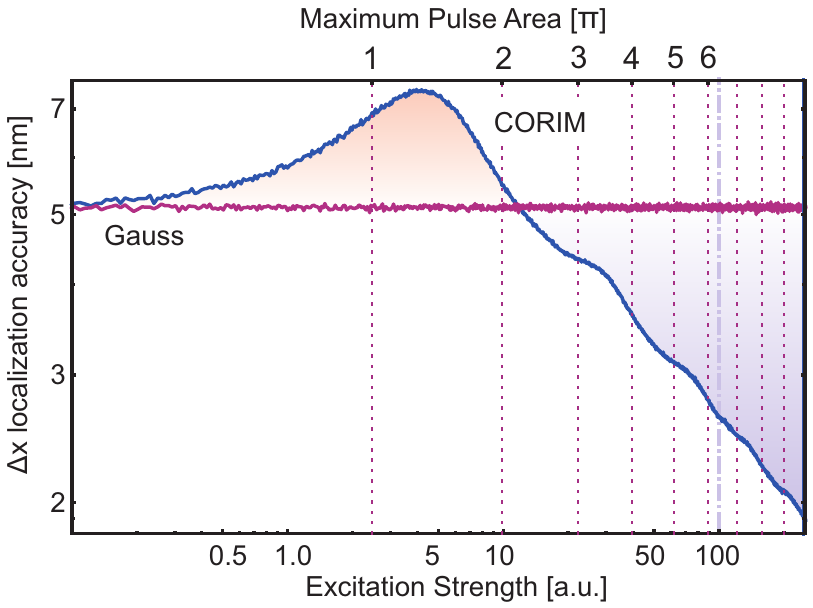}
\caption{Localization accuracy for Gaussian localization (red curve)
and for coherent Rabi imaging microscopy (blue curve). With
increasing pulse strength, the Gaussian point-spread function widens
first and reduces the localization accuracy by increasing the
effective spot size. At an intensity of about 2$\pi$, the
localization gets better than linear imaging. The oscillations for
higher field strength are explained by opening and closing the inner
spot of the point spread function. See text for details on the
excitation strength units and the details of the
simulation.\label{fig:mc01}}
\end{figure}

Linear imaging requires operation below the saturation regime
because otherwise the image of a single molecule is distorted with
respect to the expected point-spread function. In CORIM, on the
other hand, higher intensities simply result in more photo cycles.
In fact, the number of Rabi oscillations has a direct impact on the
localization accuracy. In Fig.~\ref{fig:mc01} the excitation
intensity and thus the number of oscillations are varied. This is
realized by changing the pulse area in Eqn.~\ref{eqn:nobg}. For zero
amplitude the CORIM response is equivalent to the linear imaging
such that both curves originate at one point. As the number of Rabi
oscillations increases, the localization accuracy becomes worse than
for Gaussian fitting because of a virtual broadening of the recorded
point-spread function. In other words, the same number of photons is
distributed unspecifically on a flat-top distribution. However, when
the field strength is increased further, the resulting spot is
``opened'' at the center and the localization accuracy becomes
superior to that of a Gaussian fitting. Here, the larger number of
observed Rabi oscillations helps to achieve higher spatial
frequencies, allowing for better fitting. The slight undulations of
the localization accuracy versus the field strength ($E(x_0,y_0)$)
shows that each time that the optical response opens (this occurs at
odd multipliers of $\pi$), the accuracy is improved. The slope of
the curve in Fig.~\ref{fig:mc01} yields a $1/\sqrt[4]{E(x_0,y_0)}$
behavior for the relative excitation intensity.

\begin{figure}[thb!]
\centering
\includegraphics[width=\textw]{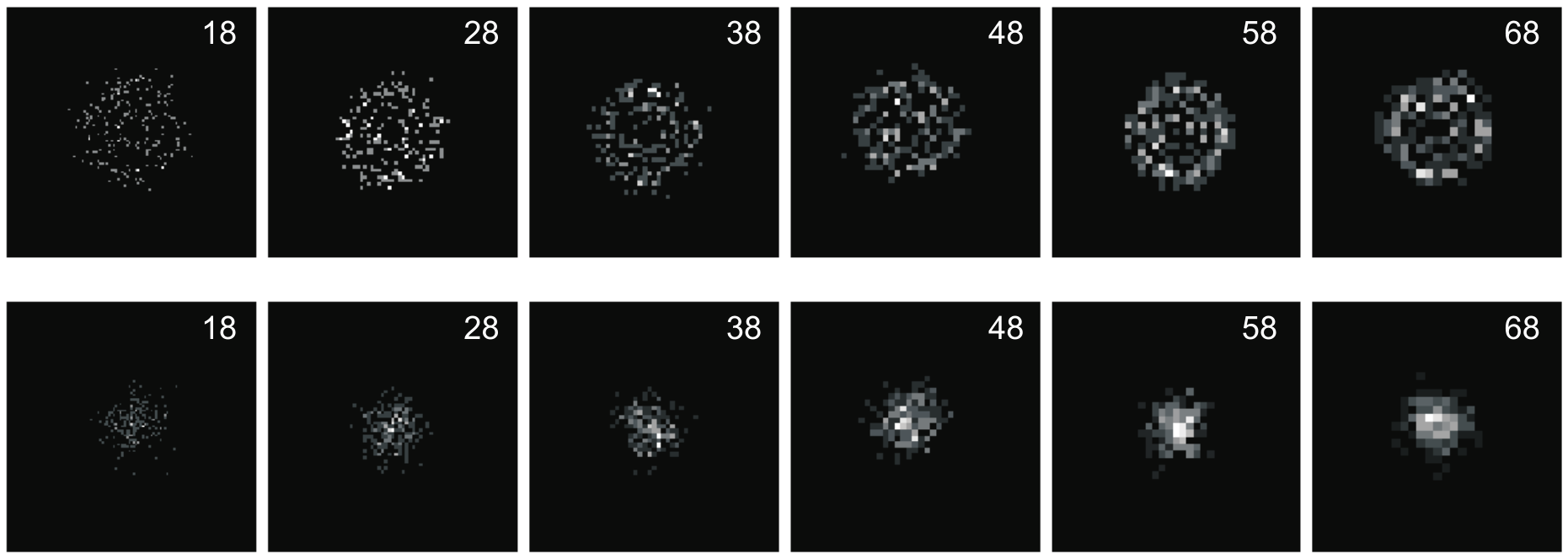}\\
\includegraphics[width=7.0cm]{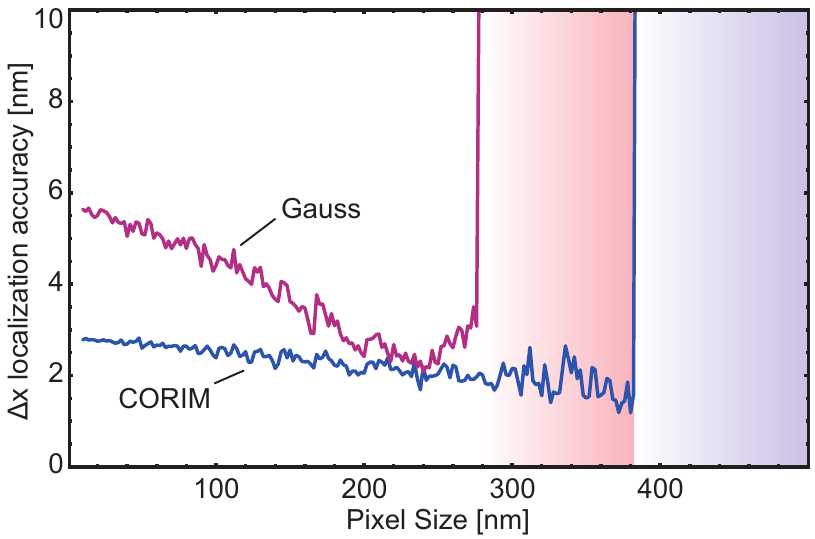}
\caption{Localization accuracy for different pixel sizes. The
coherent Rabi imaging microscopy (blue curve) has a higher stability
against fitting. Note that background noise is neglected such that
only pixilation noise is present. For details of this simulation see
text.\label{fig:mc03}}
\end{figure}

As shown by Thompson and coworkers~\cite{thompson__2002}, the
localization accuracy depends also on the pixel size. In
Fig.~\ref{fig:mc03} we present a comparison between linear imaging
and CORIM regarding the influence of the pixilation noise on the
localization accuracy. We see that in both cases the localization
procedure becomes meaningless for large pixels, however CORIM turns
out to be more tolerant in this respect.

The decisive parameters in finding the center of an emission spot
are the total number of detected photons, the pixel size, and
pixilation noise causing higher frequency components in the
image~\cite{heintzmann_natphoton_2009}. Besides these factors, any
additional source of noise introduces extra error in the
localization accuracy. One important omnipresent example stems from
the background fluorescence of the sample and other molecules in it.
In the case of pulsed excitation two other issues have to be kept in
mind. If the pulse has a finite length, there is a certain
probability that the single emitter decays and is re-excited within
the pulse duration. Furthermore, pulse fluctuations in time and
intensity wash out the modulation depth.

While we have chosen to analyze Monte Carlo simulations in this
work, a variety of other measures can be used for evaluating the
localization accuracy in microscopy~\cite{thompson__2002}. For
example, one could start with the expected point-spread function
(see Fig.~\ref{fig:fitexplanations}a), add different kinds of noise
at each pixel and analyze the localization error. A similar strategy
corresponds to the characterization of the Cramer-Rao lower bound by
calculating the Fisher information
contents~\cite{ober_biophysj_2004}. However, regardless of the exact
procedure, it is apparent that higher spatial frequencies in the
overall point-spread function provide more information and improve
the localization accuracy. In particular, the largest slope of the
point-spread function contributes most to the localization
accuracy~\cite{bobroff:1152}.

\subsection{Resolving several emitters}

Having discussed the accuracy in the localization of a single
molecule, we now explore the potential of CORIM for resolving
several close-lying emitters. Since in the experiment we only
collect red-shifted incoherent photons, we simply add the amplitudes
and do not need to include interferences~\cite{scully}.

\begin{figure}[t]
\centering
\includegraphics[width=7.0cm]{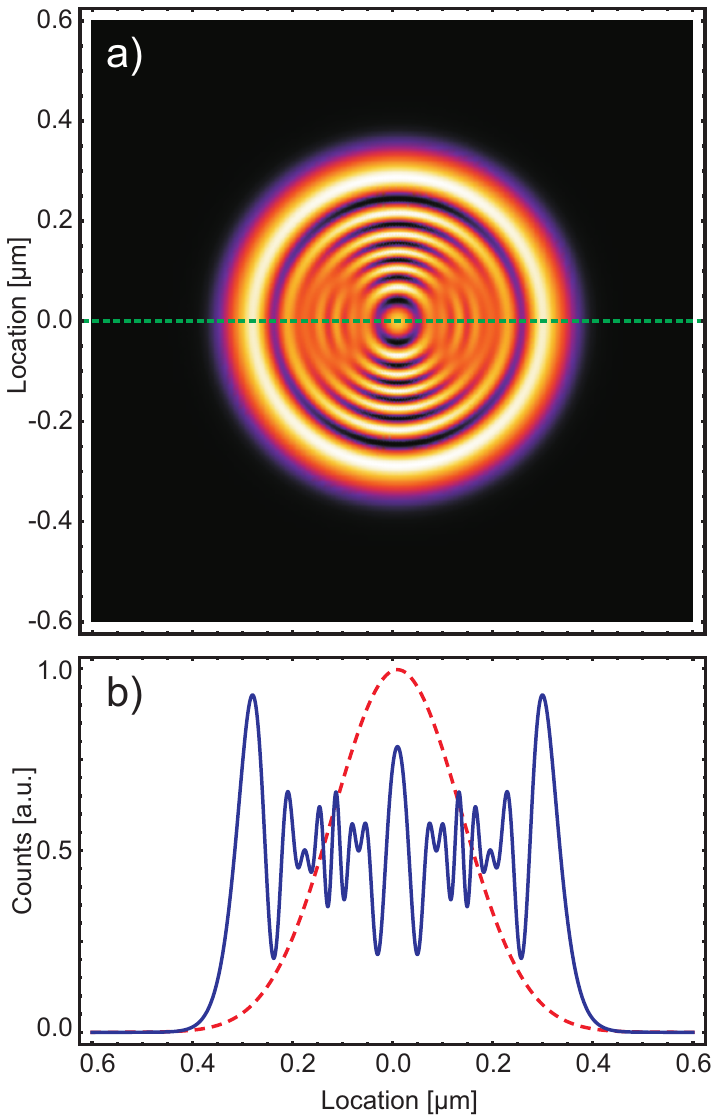}
\caption{(a) Simulated image of two molecules located at a distance
of 20~nm experiencing the same effective pulse area with the same
illumination strength. The concentric rings with sub-wavelength
features can be easily differentiated. (b) A line cut through the
central of the point-spread function. \label{fig:coupled}}
\end{figure}

Figure~\ref{fig:coupled} displays an example of two emitters
separated by 20~nm along the x-axis. On the separation axis it is
apparent that one maximum of the response coincides with the minimum
of the other emitter. The situation is analogous to magnetic
resonance imaging (MRI) or super-resolution microscopy in
inhomogeneous electric fields~\cite{hettich_science_2002}, where
externally imposed static field gradients lead to position-dependent
spectral shifts. Here the field gradient is provided by an
inhomogeneous light intensity distribution. If we consider several
concentric rings, we can assign a conservative bound to the
localization accuracy by the resulting width of the rings.

\begin{figure}[t]
\centering
\includegraphics[width=7.0cm]{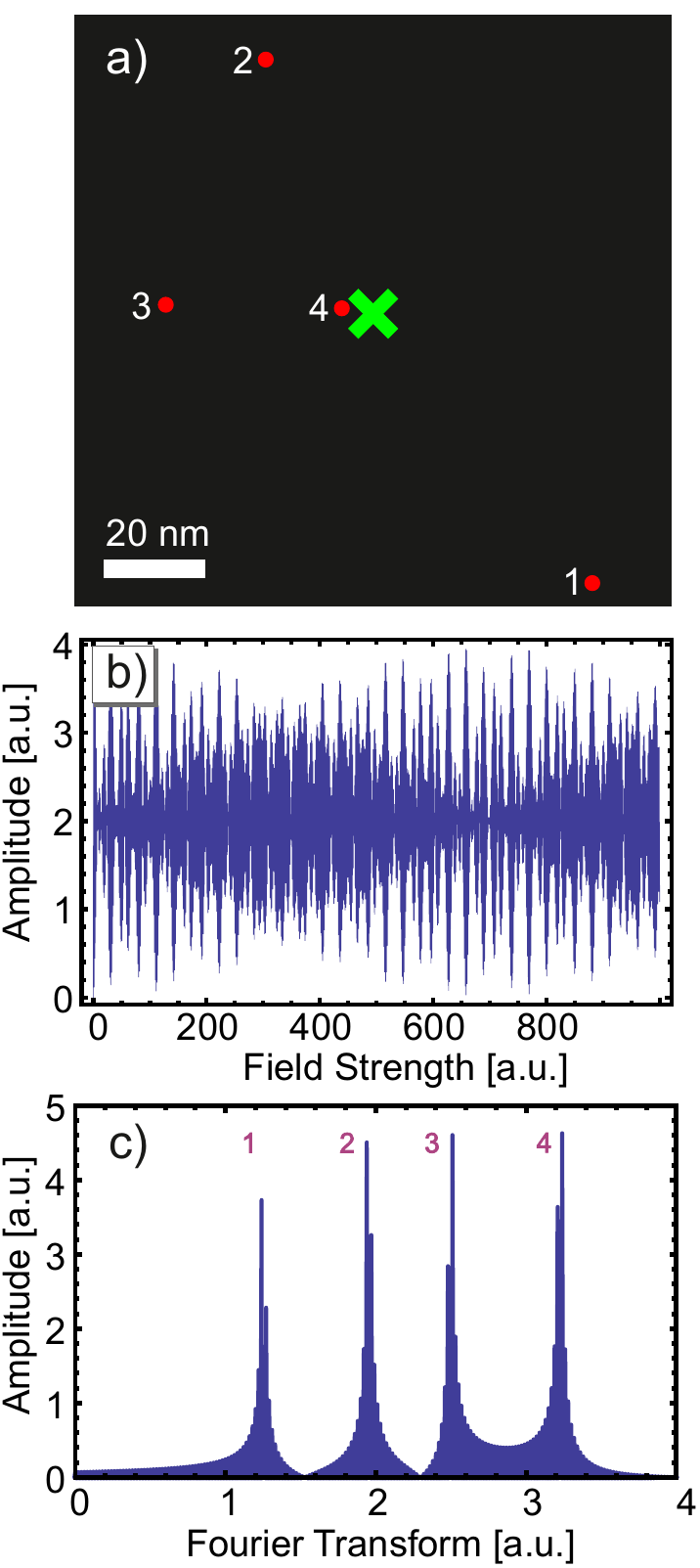}
\caption{(a) Simulated image of four molecules (red dots), located
at a small distance from the measurement point. The field dependence
measured on a central pixel (green cross) is displayed in (b). c)
The fourier transform of the Rabi oscillations reveal that although
measured only at one single pixel, all four emitters can be
determined. If the inhomogeneous broadening of the Rabi response is
negligible, the distance of all emitters to the measurement spot can
be determined.\label{fig:coupled2}}
\end{figure}

Let us start by considering emitters with negligible transition
frequency differences. As a concrete example, we assume four
close-lying emitters as depicted in Fig.~\ref{fig:coupled2}a and
recorde the Rabi response at a central pixel labeled by a green
cross. At this point the emitters feel different effective pulse
areas and thus undergo different Rabi frequencies, leading to a
complex beating pattern that results if one plots the fluorescence
signal as a function of field strength (see
Fig.~\ref{fig:coupled2}b). However, Fig.~\ref{fig:coupled2}c reveals
that the Fourier transform of this signal clearly identifies four
Rabi frequencies corresponding to the four molecules separated by a
fraction of the wavelength.

If the transition frequencies of the emitters are slightly different
(as e.g.~in the inhomogeneous band of molecules), it is still
possible to address them simultaneously by using an optical pulse
that is short enough to cover their frequency differences. If the
emitters show a considerable inhomogeneous broadening, the frequency
detuning $\Delta$ contributes to the effective Rabi frequency
according to $\mathit{\Omega}_{\rm eff} =
\sqrt{\mathit{\Omega}^2+\mathit{\Delta}^2}$. If we now scan the
focus across the sample, the Rabi response differs not only because
of the effective excitation strength, but also because of the
different spectral response. As a result, the asymmetry of the
point-spread function is larger than in the earlier case and lateral shifts
in the order of 1~nm can be resolved. This also represents a
convenient method to locate coupled molecules as it has been
observed in~\cite{hettich_science_2002}.

\subsection{Outlook}

We have introduced a method to localize emitters based on coherent
Rabi oscillations. The technique introduces higher spatial
frequencies in the recorded images, allowing for a higher
localization accuracy with fewer photons. It is possible to extend
these experiments to ambient conditions if the coherent features of
single molecules are explored further~\cite{brinks_nature_2010}.
Moreover, a single emitter can be used to sample and characterize an
optical pulse at nanoscopic scales.

The localization technique discussed here can be readily generalized
to other coherently driven processes and emitters. For example, one
can apply it to ions in a trap where numerous Rabi oscillations have
been demonstrated with a very high fidelity~\cite{roos_prl_1999}.
Furthermore, CORIM can be applied to emitters such as NV- centers in
diamond with microwave transitions of the ground state.


\subsection{Acknowledgements}
I.G. acknowledges discussions with A.~Walser and M.~Badieirostami
and continuous support from J.~Wrachtrup and F.~Jelezko. We thank
A.~Renn for valuable discussions and advice. This work was financed
by the Schweizerische Nationalfond (SNF) and the ETH Zurich
initiative for Quantum Systems for Information Technology (QSIT).


\begin{thebibliography}{21}%
\makeatletter
\providecommand \@ifxundefined [1]{%
 \@ifx{#1\undefined}
}%
\providecommand \@ifnum [1]{%
 \ifnum #1\expandafter \@firstoftwo
 \else \expandafter \@secondoftwo
 \fi
}%
\providecommand \@ifx [1]{%
 \ifx #1\expandafter \@firstoftwo
 \else \expandafter \@secondoftwo
 \fi
}%
\providecommand \natexlab [1]{#1}%
\providecommand \enquote  [1]{``#1''}%
\providecommand \bibnamefont  [1]{#1}%
\providecommand \bibfnamefont [1]{#1}%
\providecommand \citenamefont [1]{#1}%
\providecommand \href@noop [0]{\@secondoftwo}%
\providecommand \href [0]{\begingroup \@sanitize@url \@href}%
\providecommand \@href[1]{\@@startlink{#1}\@@href}%
\providecommand \@@href[1]{\endgroup#1\@@endlink}%
\providecommand \@sanitize@url [0]{\catcode `\\12\catcode `\$12\catcode
  `\&12\catcode `\#12\catcode `\^12\catcode `\_12\catcode `\%12\relax}%
\providecommand \@@startlink[1]{}%
\providecommand \@@endlink[0]{}%
\providecommand \url  [0]{\begingroup\@sanitize@url \@url }%
\providecommand \@url [1]{\endgroup\@href {#1}{\urlprefix }}%
\providecommand \urlprefix  [0]{URL }%
\providecommand \Eprint [0]{\href }%
\providecommand \doibase [0]{http://dx.doi.org/}%
\providecommand \selectlanguage [0]{\@gobble}%
\providecommand \bibinfo  [0]{\@secondoftwo}%
\providecommand \bibfield  [0]{\@secondoftwo}%
\providecommand \translation [1]{[#1]}%
\providecommand \BibitemOpen [0]{}%
\providecommand \bibitemStop [0]{}%
\providecommand \bibitemNoStop [0]{.\EOS\space}%
\providecommand \EOS [0]{\spacefactor3000\relax}%
\providecommand \BibitemShut  [1]{\csname bibitem#1\endcsname}%
\let\auto@bib@innerbib\@empty
\bibitem [{\citenamefont {Pohl}\ \emph {et~al.}(1984)\citenamefont {Pohl},
  \citenamefont {Denk},\ and\ \citenamefont {Lanz}}]{pohl:84}%
  \BibitemOpen
  \bibfield  {author} {\bibinfo {author} {\bibfnamefont {D.~W.}\ \bibnamefont
  {Pohl}}, \bibinfo {author} {\bibfnamefont {W.}~\bibnamefont {Denk}}, \ and\
  \bibinfo {author} {\bibfnamefont {M.}~\bibnamefont {Lanz}},\ }\href
  {http://link.aip.org/link/?APL/44/651/1} {\bibfield  {journal} {\bibinfo
  {journal} {Applied Physics Letters}\ }\textbf {\bibinfo {volume} {44}},\
  \bibinfo {pages} {651} (\bibinfo {year} {1984})}\BibitemShut {NoStop}%
\bibitem [{\citenamefont {Lewis}\ \emph {et~al.}(1984)\citenamefont {Lewis},
  \citenamefont {Isaacson}, \citenamefont {Harootunian},\ and\ \citenamefont
  {Muray}}]{lewis_ultramicroscopy_1984}%
  \BibitemOpen
  \bibfield  {author} {\bibinfo {author} {\bibfnamefont {A.}~\bibnamefont
  {Lewis}}, \bibinfo {author} {\bibfnamefont {M.}~\bibnamefont {Isaacson}},
  \bibinfo {author} {\bibfnamefont {A.}~\bibnamefont {Harootunian}}, \ and\
  \bibinfo {author} {\bibfnamefont {A.}~\bibnamefont {Muray}},\ }\href
  {http://www.sciencedirect.com/science/article/B6TW1-46JGN07-KS/2/6a5d3e21fdd%
ab31a83391f2ef8bc8a99} {\bibfield  {journal} {\bibinfo  {journal}
  {Ultramicroscopy}\ }\textbf {\bibinfo {volume} {13}},\ \bibinfo {pages} {227
  } (\bibinfo {year} {1984})}\BibitemShut {NoStop}%
\bibitem [{\citenamefont {Zumbusch}\ \emph {et~al.}(1999)\citenamefont
  {Zumbusch}, \citenamefont {Holtom},\ and\ \citenamefont
  {Xie}}]{PhysRevLett.82.4142}%
  \BibitemOpen
  \bibfield  {author} {\bibinfo {author} {\bibfnamefont {A.}~\bibnamefont
  {Zumbusch}}, \bibinfo {author} {\bibfnamefont {G.~R.}\ \bibnamefont
  {Holtom}}, \ and\ \bibinfo {author} {\bibfnamefont {X.~S.}\ \bibnamefont
  {Xie}},\ }\href {http://link.aps.org/doi/10.1103/PhysRevLett.82.4142}
  {\bibfield  {journal} {\bibinfo  {journal} {Physical Review Letters}\
  }\textbf {\bibinfo {volume} {82}},\ \bibinfo {pages} {4142} (\bibinfo {year}
  {1999})}\BibitemShut {NoStop}%
\bibitem [{\citenamefont {Hell}(2007)}]{hell_science_2007}%
  \BibitemOpen
  \bibfield  {author} {\bibinfo {author} {\bibfnamefont {S.~W.}\ \bibnamefont
  {Hell}},\ }\href
  {http://www.sciencemag.org/cgi/content/abstract/316/5828/1153} {\bibfield
  {journal} {\bibinfo  {journal} {Science}\ }\textbf {\bibinfo {volume}
  {316}},\ \bibinfo {pages} {1153} (\bibinfo {year} {2007})}\BibitemShut
  {NoStop}%
\bibitem [{\citenamefont {Betzig}(1995)}]{Betzig:95}%
  \BibitemOpen
  \bibfield  {author} {\bibinfo {author} {\bibfnamefont {E.}~\bibnamefont
  {Betzig}},\ }\href {http://ol.osa.org/abstract.cfm?URI=ol-20-3-237}
  {\bibfield  {journal} {\bibinfo  {journal} {Optics Letters}\ }\textbf
  {\bibinfo {volume} {20}},\ \bibinfo {pages} {237} (\bibinfo {year}
  {1995})}\BibitemShut {NoStop}%
\bibitem [{\citenamefont {Pertsinidis}\ \emph {et~al.}(2010)\citenamefont
  {Pertsinidis}, \citenamefont {Zhang},\ and\ \citenamefont
  {Chu}}]{pertsinidis_nature_2010}%
  \BibitemOpen
  \bibfield  {author} {\bibinfo {author} {\bibfnamefont {A.}~\bibnamefont
  {Pertsinidis}}, \bibinfo {author} {\bibfnamefont {Y.}~\bibnamefont {Zhang}},
  \ and\ \bibinfo {author} {\bibfnamefont {S.}~\bibnamefont {Chu}},\ }\href
  {http://dx.doi.org/10.1038/nature09163} {\bibfield  {journal} {\bibinfo
  {journal} {Nature}\ }\textbf {\bibinfo {volume} {466}},\ \bibinfo {pages}
  {647} (\bibinfo {year} {2010})}\BibitemShut {NoStop}%
\bibitem [{\citenamefont {Hettich}\ \emph {et~al.}(2002)\citenamefont
  {Hettich}, \citenamefont {Schmitt}, \citenamefont {Zitzmann}, \citenamefont
  {K\"{u}hn}, \citenamefont {Gerhardt},\ and\ \citenamefont
  {Sandoghdar}}]{hettich_science_2002}%
  \BibitemOpen
  \bibfield  {author} {\bibinfo {author} {\bibfnamefont {C.}~\bibnamefont
  {Hettich}}, \bibinfo {author} {\bibfnamefont {C.}~\bibnamefont {Schmitt}},
  \bibinfo {author} {\bibfnamefont {J.}~\bibnamefont {Zitzmann}}, \bibinfo
  {author} {\bibfnamefont {S.}~\bibnamefont {K\"{u}hn}}, \bibinfo {author}
  {\bibfnamefont {I.}~\bibnamefont {Gerhardt}}, \ and\ \bibinfo {author}
  {\bibfnamefont {V.}~\bibnamefont {Sandoghdar}},\ }\href
  {http://dx.doi.org/10.1126/science.1075606} {\bibfield  {journal} {\bibinfo
  {journal} {Science}\ }\textbf {\bibinfo {volume} {298}},\ \bibinfo {pages}
  {385} (\bibinfo {year} {2002})}\BibitemShut {NoStop}%
\bibitem [{\citenamefont {Bobroff}(1986)}]{bobroff:1152}%
  \BibitemOpen
  \bibfield  {author} {\bibinfo {author} {\bibfnamefont {N.}~\bibnamefont
  {Bobroff}},\ }\href {http://link.aip.org/link/?RSI/57/1152/1} {\bibfield
  {journal} {\bibinfo  {journal} {Review of Scientific Instruments}\ }\textbf
  {\bibinfo {volume} {57}},\ \bibinfo {pages} {1152} (\bibinfo {year}
  {1986})}\BibitemShut {NoStop}%
\bibitem [{\citenamefont {Thompson}\ \emph {et~al.}(2002)\citenamefont
  {Thompson}, \citenamefont {Larson},\ and\ \citenamefont
  {Webb}}]{thompson__2002}%
  \BibitemOpen
  \bibfield  {author} {\bibinfo {author} {\bibfnamefont {R.~E.}\ \bibnamefont
  {Thompson}}, \bibinfo {author} {\bibfnamefont {D.~R.}\ \bibnamefont
  {Larson}}, \ and\ \bibinfo {author} {\bibfnamefont {W.~W.}\ \bibnamefont
  {Webb}},\ }\href
  {http://linkinghub.elsevier.com/retrieve/pii/S000634950275618X} {\bibfield
  {journal} {\bibinfo  {journal} {Biophysical Journal}\ }\textbf {\bibinfo
  {volume} {82}},\ \bibinfo {pages} {2775} (\bibinfo {year}
  {2002})}\BibitemShut {NoStop}%
\bibitem [{\citenamefont {Betzig}\ \emph {et~al.}(2006)\citenamefont {Betzig},
  \citenamefont {Patterson}, \citenamefont {Sougrat}, \citenamefont
  {Lindwasser}, \citenamefont {Olenych}, \citenamefont {Bonifacino},
  \citenamefont {Davidson}, \citenamefont {Lippincott-Schwartz},\ and\
  \citenamefont {Hess}}]{Betzig2006}%
  \BibitemOpen
  \bibfield  {author} {\bibinfo {author} {\bibfnamefont {E.}~\bibnamefont
  {Betzig}}, \bibinfo {author} {\bibfnamefont {G.~H.}\ \bibnamefont
  {Patterson}}, \bibinfo {author} {\bibfnamefont {R.}~\bibnamefont {Sougrat}},
  \bibinfo {author} {\bibfnamefont {O.~W.}\ \bibnamefont {Lindwasser}},
  \bibinfo {author} {\bibfnamefont {S.}~\bibnamefont {Olenych}}, \bibinfo
  {author} {\bibfnamefont {J.~S.}\ \bibnamefont {Bonifacino}}, \bibinfo
  {author} {\bibfnamefont {M.~W.}\ \bibnamefont {Davidson}}, \bibinfo {author}
  {\bibfnamefont {J.}~\bibnamefont {Lippincott-Schwartz}}, \ and\ \bibinfo
  {author} {\bibfnamefont {H.~F.}\ \bibnamefont {Hess}},\ }\href
  {http://www.sciencemag.org/cgi/content/abstract/313/5793/1642} {\bibfield
  {journal} {\bibinfo  {journal} {Science}\ }\textbf {\bibinfo {volume}
  {313}},\ \bibinfo {pages} {1642} (\bibinfo {year} {2006})}\BibitemShut
  {NoStop}%
\bibitem [{\citenamefont {Rust}\ \emph {et~al.}(2006)\citenamefont {Rust},
  \citenamefont {Bates},\ and\ \citenamefont {Zhuang}}]{rust_natmeth_2006}%
  \BibitemOpen
  \bibfield  {author} {\bibinfo {author} {\bibfnamefont {M.~J.}\ \bibnamefont
  {Rust}}, \bibinfo {author} {\bibfnamefont {M.}~\bibnamefont {Bates}}, \ and\
  \bibinfo {author} {\bibfnamefont {X.}~\bibnamefont {Zhuang}},\ }\href
  {http://dx.doi.org/10.1038/nmeth929} {\bibfield  {journal} {\bibinfo
  {journal} {Nature Methods}\ }\textbf {\bibinfo {volume} {3}},\ \bibinfo
  {pages} {793} (\bibinfo {year} {2006})}\BibitemShut {NoStop}%
\bibitem [{\citenamefont {Gerhardt}\ \emph {et~al.}(2009)\citenamefont
  {Gerhardt}, \citenamefont {Wrigge}, \citenamefont {Zumofen}, \citenamefont
  {Hwang}, \citenamefont {Renn},\ and\ \citenamefont
  {Sandoghdar}}]{gerhardt_pra_2009}%
  \BibitemOpen
  \bibfield  {author} {\bibinfo {author} {\bibfnamefont {I.}~\bibnamefont
  {Gerhardt}}, \bibinfo {author} {\bibfnamefont {G.}~\bibnamefont {Wrigge}},
  \bibinfo {author} {\bibfnamefont {G.}~\bibnamefont {Zumofen}}, \bibinfo
  {author} {\bibfnamefont {J.}~\bibnamefont {Hwang}}, \bibinfo {author}
  {\bibfnamefont {A.}~\bibnamefont {Renn}}, \ and\ \bibinfo {author}
  {\bibfnamefont {V.}~\bibnamefont {Sandoghdar}},\ }\href
  {http://link.aps.org/doi/10.1103/PhysRevA.79.011402} {\bibfield  {journal}
  {\bibinfo  {journal} {Physical Review A}\ }\textbf {\bibinfo {volume} {79}},\
  \bibinfo {pages} {011402} (\bibinfo {year} {2009})}\BibitemShut {NoStop}%
\bibitem [{\citenamefont {Orrit}\ and\ \citenamefont
  {Bernard}(1990)}]{Orrit:90}%
  \BibitemOpen
  \bibfield  {author} {\bibinfo {author} {\bibfnamefont {M.}~\bibnamefont
  {Orrit}}\ and\ \bibinfo {author} {\bibfnamefont {J.}~\bibnamefont
  {Bernard}},\ }\href {http://dx.doi.org/10.1103/PhysRevLett.65.2716}
  {\bibfield  {journal} {\bibinfo  {journal} {Physical Review Letters}\
  }\textbf {\bibinfo {volume} {65}},\ \bibinfo {pages} {2716} (\bibinfo {year}
  {1990})}\BibitemShut {NoStop}%
\bibitem [{\citenamefont {Wrigge}\ \emph {et~al.}(2008)\citenamefont {Wrigge},
  \citenamefont {Gerhardt}, \citenamefont {Hwang}, \citenamefont {Zumofen},\
  and\ \citenamefont {Sandoghdar}}]{Wrigge:08}%
  \BibitemOpen
  \bibfield  {author} {\bibinfo {author} {\bibfnamefont {G.}~\bibnamefont
  {Wrigge}}, \bibinfo {author} {\bibfnamefont {I.}~\bibnamefont {Gerhardt}},
  \bibinfo {author} {\bibfnamefont {J.}~\bibnamefont {Hwang}}, \bibinfo
  {author} {\bibfnamefont {G.}~\bibnamefont {Zumofen}}, \ and\ \bibinfo
  {author} {\bibfnamefont {V.}~\bibnamefont {Sandoghdar}},\ }\href
  {http://dx.doi.org/10.1038/nphys812} {\bibfield  {journal} {\bibinfo
  {journal} {Nature Physics}\ }\textbf {\bibinfo {volume} {4}},\ \bibinfo
  {pages} {60} (\bibinfo {year} {2008})}\BibitemShut {NoStop}%
\bibitem [{\citenamefont {Ippolito}\ \emph {et~al.}(2005)\citenamefont
  {Ippolito}, \citenamefont {Goldberg},\ and\ \citenamefont
  {\"{U}nl\"{u}}}]{ippolito:053105}%
  \BibitemOpen
  \bibfield  {author} {\bibinfo {author} {\bibfnamefont {S.~B.}\ \bibnamefont
  {Ippolito}}, \bibinfo {author} {\bibfnamefont {B.~B.}\ \bibnamefont
  {Goldberg}}, \ and\ \bibinfo {author} {\bibfnamefont {M.~S.}\ \bibnamefont
  {\"{U}nl\"{u}}},\ }\href {http://link.aip.org/link/?JAP/97/053105/1}
  {\bibfield  {journal} {\bibinfo  {journal} {Journal of Applied Physics}\
  }\textbf {\bibinfo {volume} {97}},\ \bibinfo {eid} {053105} (\bibinfo {year}
  {2005})}\BibitemShut {NoStop}%
\bibitem [{\citenamefont {Bloe\ss}\ \emph {et~al.}(2001)\citenamefont
  {Bloe\ss}, \citenamefont {Durand}, \citenamefont {Matsushita}, \citenamefont
  {{v}an~{d}er Meer}, \citenamefont {Brakenhoff},\ and\ \citenamefont
  {Schmidt}}]{bloess01}%
  \BibitemOpen
  \bibfield  {author} {\bibinfo {author} {\bibfnamefont {A.}~\bibnamefont
  {Bloe\ss}}, \bibinfo {author} {\bibfnamefont {Y.}~\bibnamefont {Durand}},
  \bibinfo {author} {\bibfnamefont {M.}~\bibnamefont {Matsushita}}, \bibinfo
  {author} {\bibfnamefont {H.}~\bibnamefont {{v}an~{d}er Meer}}, \bibinfo
  {author} {\bibfnamefont {G.~J.}\ \bibnamefont {Brakenhoff}}, \ and\ \bibinfo
  {author} {\bibfnamefont {J.}~\bibnamefont {Schmidt}},\ }\href
  {http://dx.doi.org/10.1046/j.0022-2720.2001.00971.x} {\bibfield  {journal}
  {\bibinfo  {journal} {Journal of Microscopy}\ }\textbf {\bibinfo {volume}
  {205}},\ \bibinfo {pages} {76} (\bibinfo {year} {2001})}\BibitemShut
  {NoStop}%
\bibitem [{\citenamefont {Heintzmann}\ and\ \citenamefont
  {Gustafsson}(2009)}]{heintzmann_natphoton_2009}%
  \BibitemOpen
  \bibfield  {author} {\bibinfo {author} {\bibfnamefont {R.}~\bibnamefont
  {Heintzmann}}\ and\ \bibinfo {author} {\bibfnamefont {M.~G.~L.}\ \bibnamefont
  {Gustafsson}},\ }\href {http://dx.doi.org/10.1038/nphoton.2009.102}
  {\bibfield  {journal} {\bibinfo  {journal} {Nature Photonics}\ }\textbf
  {\bibinfo {volume} {3}},\ \bibinfo {pages} {362} (\bibinfo {year}
  {2009})}\BibitemShut {NoStop}%
\bibitem [{\citenamefont {Ober}\ \emph {et~al.}(2004)\citenamefont {Ober},
  \citenamefont {Ram},\ and\ \citenamefont {Ward}}]{ober_biophysj_2004}%
  \BibitemOpen
  \bibfield  {author} {\bibinfo {author} {\bibfnamefont {R.~J.}\ \bibnamefont
  {Ober}}, \bibinfo {author} {\bibfnamefont {S.}~\bibnamefont {Ram}}, \ and\
  \bibinfo {author} {\bibfnamefont {E.~S.}\ \bibnamefont {Ward}},\ }\href
  {http://dx.doi.org/10.1016/S0006-3495(04)74193-4} {\bibfield  {journal}
  {\bibinfo  {journal} {Biophysical Journal}\ }\textbf {\bibinfo {volume}
  {86}},\ \bibinfo {pages} {1185} (\bibinfo {year} {2004})}\BibitemShut
  {NoStop}%
\bibitem [{\citenamefont {Scully}\ and\ \citenamefont
  {Zubairy}(1997)}]{scully}%
  \BibitemOpen
  \bibfield  {author} {\bibinfo {author} {\bibfnamefont {M.~O.}\ \bibnamefont
  {Scully}}\ and\ \bibinfo {author} {\bibfnamefont {M.~S.}\ \bibnamefont
  {Zubairy}},\ }\href@noop {} {\emph {\bibinfo {title} {Quantum Optics}}},\
  \bibinfo {edition} {1st}\ ed.\ (\bibinfo  {publisher} {Cambridge University
  Press},\ \bibinfo {year} {1997})\BibitemShut {NoStop}%
\bibitem [{\citenamefont {Brinks}\ \emph {et~al.}(2010)\citenamefont {Brinks},
  \citenamefont {Stefani}, \citenamefont {Kulzer}, \citenamefont {Hildner},
  \citenamefont {Taminiau}, \citenamefont {Avlasevich}, \citenamefont
  {M\"ullen},\ and\ \citenamefont {van Hulst}}]{brinks_nature_2010}%
  \BibitemOpen
  \bibfield  {author} {\bibinfo {author} {\bibfnamefont {D.}~\bibnamefont
  {Brinks}}, \bibinfo {author} {\bibfnamefont {F.~D.}\ \bibnamefont {Stefani}},
  \bibinfo {author} {\bibfnamefont {F.}~\bibnamefont {Kulzer}}, \bibinfo
  {author} {\bibfnamefont {R.}~\bibnamefont {Hildner}}, \bibinfo {author}
  {\bibfnamefont {T.~H.}\ \bibnamefont {Taminiau}}, \bibinfo {author}
  {\bibfnamefont {Y.}~\bibnamefont {Avlasevich}}, \bibinfo {author}
  {\bibfnamefont {K.}~\bibnamefont {M\"ullen}}, \ and\ \bibinfo {author}
  {\bibfnamefont {N.~F.}\ \bibnamefont {van Hulst}},\ }\href
  {http://dx.doi.org/10.1038/nature09110} {\bibfield  {journal} {\bibinfo
  {journal} {Nature}\ }\textbf {\bibinfo {volume} {465}},\ \bibinfo {pages}
  {905} (\bibinfo {year} {2010})}\BibitemShut {NoStop}%
\bibitem [{\citenamefont {Roos}\ \emph {et~al.}(1999)\citenamefont {Roos},
  \citenamefont {Zeiger}, \citenamefont {Rohde}, \citenamefont {N\"agerl},
  \citenamefont {Eschner}, \citenamefont {Leibfried}, \citenamefont
  {Schmidt-Kaler},\ and\ \citenamefont {Blatt}}]{roos_prl_1999}%
  \BibitemOpen
  \bibfield  {author} {\bibinfo {author} {\bibfnamefont {C.}~\bibnamefont
  {Roos}}, \bibinfo {author} {\bibfnamefont {T.}~\bibnamefont {Zeiger}},
  \bibinfo {author} {\bibfnamefont {H.}~\bibnamefont {Rohde}}, \bibinfo
  {author} {\bibfnamefont {H.~C.}\ \bibnamefont {N\"agerl}}, \bibinfo {author}
  {\bibfnamefont {J.}~\bibnamefont {Eschner}}, \bibinfo {author} {\bibfnamefont
  {D.}~\bibnamefont {Leibfried}}, \bibinfo {author} {\bibfnamefont
  {F.}~\bibnamefont {Schmidt-Kaler}}, \ and\ \bibinfo {author} {\bibfnamefont
  {R.}~\bibnamefont {Blatt}},\ }\href
  {http://link.aps.org/abstract/PRL/v83/p4713} {\bibfield  {journal} {\bibinfo
  {journal} {Physical Review Letters}\ }\textbf {\bibinfo {volume} {83}},\
  \bibinfo {pages} {4713} (\bibinfo {year} {1999})}\BibitemShut {NoStop}%
\end{thebibliography}
%

\end{document}